\documentclass[aps,prd,superscriptaddress, 10pt,twocolumn]{revtex4}
\usepackage{graphicx}
\usepackage{epstopdf}
\usepackage{amsmath}
\usepackage{amsfonts}
\usepackage{amssymb}
\usepackage{latexsym}
\usepackage{hyperref}
\usepackage[english]{babel}
\usepackage[utf8]{inputenc}
\usepackage[colorinlistoftodos]{todonotes}
\usepackage{xcolor}
\usepackage{slashed}
\usepackage{bm}  
\usepackage[normalem]{ulem} 
\usepackage{amsmath,mathtools} 
\usepackage{caption}
\usepackage{subcaption}
\usepackage{float}
\useunder{\uline}{\ul}{}
\usepackage{hyperref}
\begin{document}
\title{Elastic light-by-light scattering in a nonminimal Lorentz violation scenario}

\author{Y.M.P. Gomes}\email{ymuller@cbpf.br}
\affiliation{Centro Brasileiro de Pesquisas F\'{i}sicas (CBPF), Rua Dr Xavier Sigaud 150, Urca, Rio de Janeiro, Brazil, CEP 22290-180}
\author{J.T. Guaitolini Junior}\email{jguaitolini@ifes.edu.br}
\affiliation{Centro Brasileiro de Pesquisas F\'{i}sicas (CBPF), Rua Dr Xavier Sigaud 150, Urca, Rio de Janeiro, Brazil, CEP 22290-180}
\affiliation{Instituto Federal do Esp\'{i}rito Santo (IFES) - Campus Vit\'{o}ria, Av. Vit\'{o}ria 1729, Jucutuquara, Vit\'{o}ria, ES, Brazil, CEP 29040-780}

\begin{abstract}
Over the past years, a Lorentz symmetry violation scheme has been implemented in the Standard Model of Particle Physics, in an attempt to explain important open problems. In connection with this work, we determine the effect of two Lorentz-violating nonminimal couplings on the differential cross section for elastic photon-photon scattering, both in the high- and low-energy regimes. Novel characteristics are pointed out; in particular, a periodic pattern on the azimuthal angle and an energy-independent contribution  which are not present in Quantum Electrodynamics calculations.   
\end{abstract}
\maketitle

\section{Introduction}

\paragraph*{}
The Standard Model of Particle Physics (SM) has been successful over the past years in its attempt to describe high-energy physical phenomena. The measurements carried out in particle accelerators have shown a remarkable agreement between theory and experiment. However, despite this success, there are still open issues such as the hierarchy, the strong CP and the cosmological constant problems~\cite{CP,CC}, for which the SM offers no satisfactory explanation. Therefore, extensions of the SM have been proposed with the goal to clarify these phenomena, including extra symmetries, as in Grand Unification models, or supersymmetric extensions~\cite{GUSM,SSSM}.

\paragraph*{}
An alternative approach is the so-called Standard-Model Extension (SME)~\cite{kostint, kostint1}, which proposes a completion of the usual SM by including new Lorentz-violating (LV) interactions. The main idea is that vector or tensor fields could acquire non-zero vacuum expectations values, implying that the Lorentz symmetry should be broken by these vacuum expectation values. Possible effects in the SM processes have already been analyzed by considering both minimal and nonminimal LV couplings \cite{Mewes, LSVbhabha, LSVsaladirac, YM, tables}. Up to now, the bounds on the LV parameters are strong and corroborate the assumed validity of Lorentz symmetry, at least at the present cosmological time.

\paragraph*{}
In this work, we shall analyze the contribution of two nonminimal LV couplings in Quantum Electrodynamics (QED), specifically to the elastic scattering between two photons, also called elastic light-by-light (LbyL) scattering.

\paragraph*{}
This paper is organized as follows: in Section II, we discuss the photon-photon scattering in more detail; in Section III, we introduce the nonminimal couplings which are analyzed in this work and, in Section IV, we calculate the contribution of the nonminimal LV couplings to elastic LbyL scattering. Finally, in Section V, we discuss our results.

\section{Elastic light-by-light scattering}

\paragraph*{}
The interest in the nature of light and associated phenomena have always been present in Physics. For example, names like René Descartes, Isaac Newton, Robert Hooke, Christiaan Huyghens created wave and corpuscular models for light, and Quantum Mechanics brought with it the concept of wave-particle duality, reconciling the two points of view. However, over the past decades, new issues have continuously emerged.

Maxwell's equations of classical electrodynamics had the inclusion of light in the theory as a great success, but the linearity of the equations forbids the existence of processes allowed from the quantum point of view. As early as 1933, the concern with the properties of the quantum vacuum and the interaction between light quanta  \cite{halpern33} opened up the era of nonlinear electromagnetism. Theoretical work on nonlinear electrodynamics first appeared in the 1930s with Halpern, Born, Infeld, Euler and Heisenberg~\cite{born34, BornInf34, EulHei36, Euler36} and continued in subsequent years~\cite{KarNew50, KarNew51}. The implemented corrections resulted in the possibility of scattering between two photons by means of vacuum fluctuations~\cite{EulHei36, KarNew51} and allowed the calculation of the associated cross section~\cite{dTollis1, dTollis2, dTollis3}.

\begin{figure}[htb]
	\begin{center}
		\leavevmode
		\includegraphics[width=0.4\textwidth]{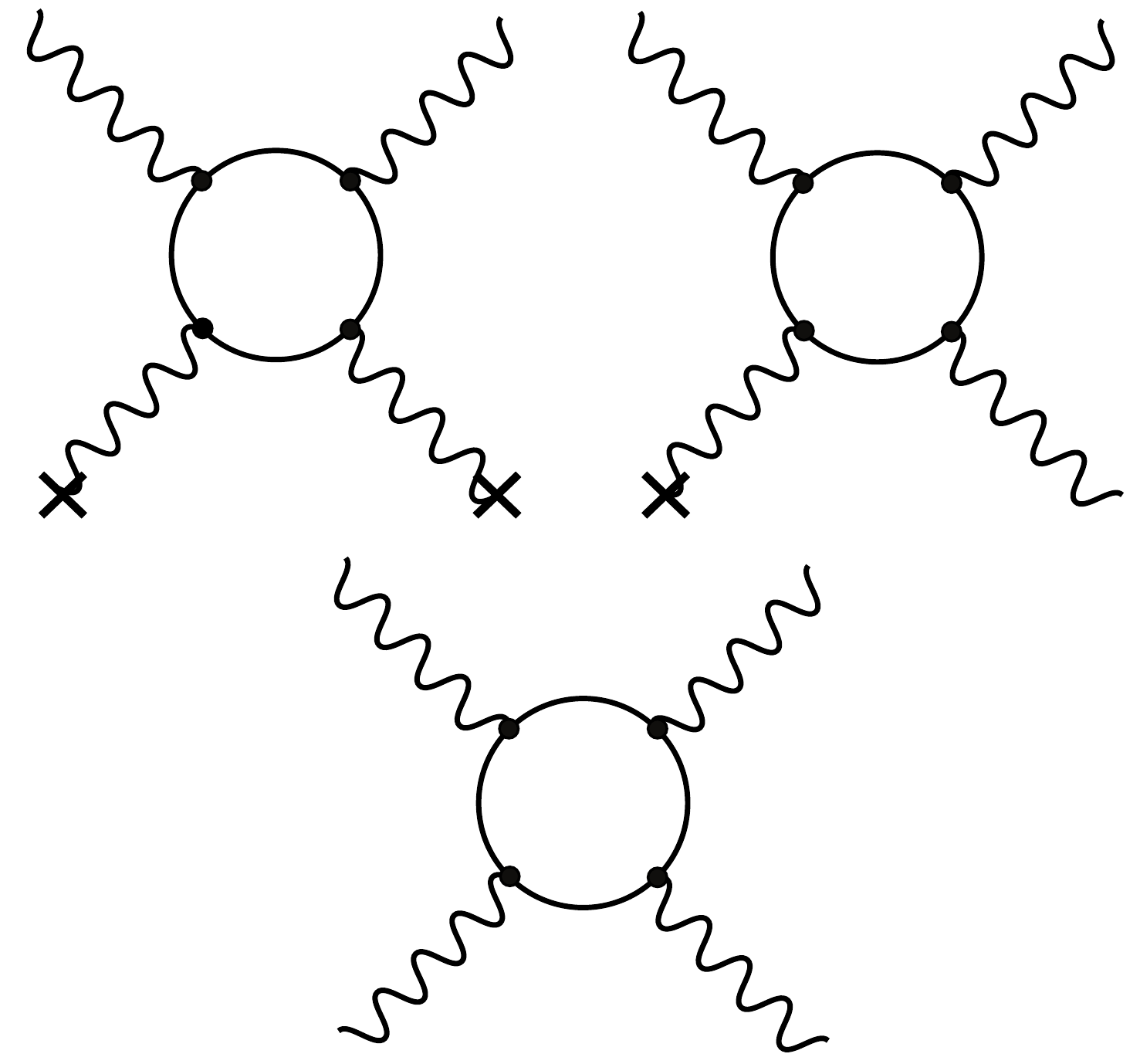}
	\end{center}
	\caption[Possible interaction between photons]{Illustration of 1-loop photon-photon interactions:  Delbr\"{u}ck scattering (top-left), photon splitting (top-right) and elastic LbyL scattering (bottom). The crosses indicate external fields, such as Coulomb or magnetic fields, and the blob represents the vertex.}
	\label{tiposfoton}
\end{figure}

\paragraph*{}
Nonlinear phenomena such as the scattering of a photon in a Coulomb field~\cite{delbruck33} or the splitting of a photon in the presence of an external field~\cite{baier74} were studied and were already observed experimentally~\cite{jarlskog73,papatzacos75,schumacher75,johannessen80,akhmadaliev2002,lee2003}. In fact, photon splitting has also been studied in the context of LV in the minimal SME~\cite{KostSplitt2003}. For elastic LbyL scattering, experimental evidence has also been observed, but due to its small cross section this happened only very recently~\cite{LbyLCERN2017, LbyLCMS2018}. These nonlinear processes are represented in the lowest order by 1-loop diagrams with four external photonic legs; but, in some cases, we replace real photons by a line representing an external field, as the examples in Fig.~\eqref{tiposfoton} show.

\paragraph*{}
From the six Feynman diagrams associated with elastic LbyL scattering obtained by the different combinations of the photonic legs, we can calculate the scattering amplitudes and the differential cross section for unpolarized photons. The loop can contain different kinds of virtual charged particles (quarks, leptons, W$^\pm$~\cite{loopW}), depending on the energy available in the experiment.

\paragraph*{}
Considering only QED vertices, in a low-energy regime ($\omega \ll m$), the differential cross section of the process is given by~\cite{dTollis1, dTollis2,delbruck33}
\begin{equation}
\label{eqLowRegime}
\frac{d \sigma}{d \Omega}^{\!\! \gamma\gamma } = \frac{139\alpha^4}{(180\pi)^2 m^2} \left( \frac{\omega}{m} \right)^6 (3 + \cos^2 \theta)^2 \, ,
\end{equation}
whereas in the ultrarelativistic case \cite{berestetskiibookedq}
\begin{equation}
\label{eqHighRegime}
\frac{d \sigma}{d \Omega}^{\!\! \gamma\gamma } = \frac{\alpha^4}{\pi^2 \omega^2} \log^4 \frac{1}{\theta} \, ,
\end{equation}
that is suitable for small scattering angles ($ m/ \omega \! \ll \! \theta \! \ll \! 1$). In both results, we are using natural units ($\hbar=c=1$) and $m$ stands for the electron mass.

\paragraph*{}
The viability of the direct detection of elastic LbyL scattering with laser beams at SLAC was already discussed in the 1980s~\cite{LbyLSLAC82}. Investigations of LbyL scattering on the visible scale, with high intensity pulsed lasers~\cite{moulin1996} obtained a first upper limit on $\sigma^{\!\! \gamma\gamma } = 10^{-39}$ cm$^2$, with $95 \%$ confidence level.  Subsequently, when upgrading the measurement with a third laser beam, the limit was improved to  $1.5 \cdot 10^{-48}$ cm$^2$~\cite{bernard2000}. In this last situation, under low-energy conditions of the experiment, the result was $18$ orders of magnitude from the result estimated by QED ($7.3 \cdot 10^{-66}$ cm$^2$). Another possibility was to use X-ray pulses (high-energy limit), avoiding the QED cross section suppressed by the sixth power of the ratio $\omega/m$~\cite{xRay1,xRay2}. In such X-ray experiments, QED cross section was estimated in  $2.5 \cdot 10^{-43}$ cm$^2$, the upper limit found was $1.9 \cdot 10^{-23}$ cm$^2$ \cite{xRay2}.

\paragraph*{}
An alternative way to inspect LbyL interactions is by using ultraperipheral heavy-ion collisions~\cite{2013-16-LbyL, gawenda2016, szczurek, klein2017}. Experiments of ultraperipheral Pb-Pb collisions performed by the ATLAS experiment with energies of $5.02$ TeV presented evidence of elastic LbyL scattering, and thirteen candidate events were observed~\cite{LbyLCERN2017}. More recently, fourteen candidate events passing all selection requirements have been reported by the CMS Collaboration~\cite{LbyLCMS2018}. However, since the number of events associated with this phenomenon was small, the analysis is still limited. Although there are already high-precision experiments to explore properties of fundamental particles, such as those based on Penning traps~\cite{dehmelt68, ding2016}, we still consider valid to analyze possible signs of LV in LbyL scattering. New LHC updates, for example, should improve the data availability, opening up new possibilities for studying physics beyond the Standard Model (BSM).

\paragraph*{}
Elastic LbyL scattering is used to constrain nonlinear corrections to Maxwell electrodynamics~\cite{Medeiros2018} and it can also provide contribution to the anomalous magnetic moment of the muon \cite{ gmuon1, gmuon2}, including chiral theories~\cite{gmuonchiral} and quantum chromodynamics (QCD) calculations, both for holographic models~\cite{gmuonQCD1} and lattice QCD~\cite{gmuonQCD2}. Furthermore, the signal of the vacuum processes are affected if new particles are coupled to photons. Searches for physics BSM may include, for instance, axion-like particles~\cite{axion0,axion1,axion2} and supersymmetric QED~\cite{susyQED}. Our proposal also aims to investigate possible evidence of physics BSM, more specifically the search for LV by nonminimal couplings that modify the interaction vertex, as already developed for other QED processes~\cite{LSVbhabha,LSVsaladirac}.

\section{Nonminimal Couplings}

\paragraph*{}
In our paper, we shall introduce a nondynamical background 4-vector, $\xi^\mu$, coupled nonminimally to the electromagnetic field strength, $F_{\mu\nu}$ (or its dual, $\tilde{F}_{\mu\nu}$) and the electromagnetic current. This fixed 4-vector will then be responsible for the  breaking of Lorentz symmetry, once it selects a privileged direction in spacetime.

\paragraph*{}
In an extended version of QED with the coupling between $\xi^\mu$ and $F_{\mu\nu}$, our Lagrangian presents the following dimension-5 term
\begin{equation}\label{nonm1}
\mathcal{L}_{\textmd{LV}} =  \xi^\mu \bar{\psi} \gamma^\nu \psi F_{\mu \nu} \, ,
\end{equation}
where $\xi^\mu$ is constant, has canonical dimension of inverse mass and is expected as a very small intensity term; $\mathcal{L}_{\textmd{LV}} $ is CPT-even whenever $\xi_\mu$ transforms under $T$-symmetry as $\xi_\mu=( \xi_0, \boldsymbol{\xi} ) \rightarrow \xi_\mu ' = (- \xi_0, \boldsymbol{\xi} )$. On the other hand,  $\mathcal{L}_{\textmd{LV}} $ would be CPT-odd if $\xi_\mu=( \xi_0, \boldsymbol{\xi} ) \rightarrow \xi_\mu ' = ( \xi_0, -\boldsymbol{\xi} )$. It describes a sort of transition electric dipole moment, which connects the relativistically dominant component of the fermion bispinor with the weak relativistic component. In terms of a standard notation among experimentalists and many theorists who deal with Lagrangian densities including dimension-5 operators \cite{ding2016, kost2018}, our background vector can be written as

\begin{equation}
\xi^{\mu}=-\frac{1}{3}a_{F~~~~\alpha}^{(5)\alpha \mu} \, .
\end{equation}

Differently from minimal couplings (e.g., Carrol-Field-Jackiw (CFJ)~\cite{CFJ}), in this case the structure of the fermionic current will be modified. So, with this new coupling, together with the QED Lagrangian
\begin{equation}
\mathcal{L}_{\textmd{QED}} = -\frac{1}{4}F^{\mu \nu}F_{\mu \nu} +  \bar{\psi} (i \slashed{D} - m) \psi~~,
\end{equation}
where $\slashed D = \gamma^\mu (\partial_\mu + i e A_\mu)$, the modified inhomogeneous Maxwell equations read as below
\begin{eqnarray}
\label{maxwell1}
\nonumber
\partial_{\mu}F^{\mu \nu} &=& e \bar{\psi} \gamma^\nu \psi + \xi^{[\nu} \partial_\mu (\bar{\psi}\gamma^{\mu]} \psi) \\
&=&(e \delta_\mu^\nu + \xi^{\nu} \partial_{\mu} - \xi_\mu \partial^\nu) (\bar{\psi}\gamma^{\mu} \psi) 
\end{eqnarray}
and 
\begin{equation}
\label{maxwell2}
(i \gamma^\mu \partial_\mu + m + e \gamma^\mu A_\mu + \gamma^\mu \xi^\nu F_{\mu \nu}) \psi = 0 \, .
\end{equation}

\paragraph*{}
Equation~\eqref{maxwell1} represents the new Maxwell equations with sources and Eq. \eqref{maxwell2} is the modified Dirac equation. By means of Eq. \eqref{maxwell1}, we can show that, in addition to the QED electric charge $Q$, a modified charge $Q'$ will be conserved. We have $ \partial_\nu {J'}^\nu = 0$, where ${J'}^\nu = (e \delta_\mu^\nu + \xi^{\nu} \partial_{\mu} - \xi_\mu \partial^\nu) (\bar{\psi}\gamma^{\mu} \psi) $ and
\begin{equation}
Q' = \int d^3x {J'}^0 =  Q +\partial_t \big[ \! \int \! d^3 x \, (\boldsymbol{\xi} \! \cdot \! \boldsymbol{J}) \big] ~, 
\end{equation} 
where $Q =  e\int d^3x \psi^\dagger \psi$ and $\boldsymbol{J}^i = e\psi^\dagger \gamma^0 \gamma^i \psi$. Though the charge is defined for the free particle, the introduction of $\mathcal{L}_{\textmd{LV}}$ yields a new conserved 4-current, $J'_\mu$. The electric charge, $Q$, conservation is not affected by the extra contribution. From a quantum-mechanical point of view, the coupling in Eq.~\eqref{nonm1} modifies the QED electron-photon vertex, which now reads
\begin{equation}
\Gamma^\mu = e \gamma^\mu - i \slashed{q}\xi^\mu + i (\xi \cdot q) \gamma^\mu \, .
\end{equation}

\paragraph*{}
As previously mentioned, we can consider other non-minimal interaction by coupling our background 4-vector to the dual of the electromagnetic field strength~\cite{belich2006}. The new LV term modifying the standard QED Lagrangian is then 
\begin{equation}
\label{nonm22}
\mathcal{L}_{\textmd{LV}} = \tilde{\xi}^\mu \bar{\psi} \gamma^\nu \psi \tilde{F}_{\mu \nu} \, ,
\end{equation}
where $\tilde{F}_{\mu \nu} = \frac{1}{2}\varepsilon_{\mu \nu \alpha \beta} F^{\alpha \beta}$ and $\tilde{\xi}^\mu$ has the same characteristics defined for $\xi^{\mu}$. Differently from the term in Eq. \eqref{nonm1}, this term does not contribute to CP-violating decays, but it yields a sort of transition magnetic dipole moment for the fermions. Again, using the standard notation previously mentioned, we have
\begin{equation}
\tilde{\xi}^{\mu} = \frac{1}{6}\varepsilon^{\mu}_{~\nu \alpha \beta} a_{F}^{(5)\nu \alpha \beta} \, .
\end{equation}

The modified Maxwell equations becomes
\begin{eqnarray}\nonumber
\partial_{\mu}F^{\mu \nu} &=& e \bar{\psi} \gamma^\nu \psi +  \varepsilon^{\mu \nu}_{~~\alpha \beta} \tilde{\xi}^{\alpha} \partial_\mu \bar{\psi}\gamma^{\beta} \psi  \\
&=& (e \delta^\nu_\alpha -  \varepsilon^\nu_{~  \alpha \mu \beta}\tilde{\xi}^\mu \partial^\beta) \bar{\psi}\gamma^{\alpha} \psi 
\end{eqnarray}
and 
\begin{equation}
(i \gamma^\mu \partial_\mu + m + e \gamma^\mu A_\mu + \gamma^\mu \tilde{\xi}^\nu \tilde{F}_{\mu \nu}) \psi = 0 \, ;
\end{equation}
an important detail here is that the QED conserved charge does not change, i.e., $Q' = Q = e\int d^3x\psi^\dagger \psi$. In this case, the extension of the usual QED vertex is written as
\begin{equation}
\label{nonm2}
\Gamma^\mu =  e\gamma^\mu - \varepsilon^\mu_{~  \alpha \nu \beta}\gamma^\alpha \tilde{\xi}^\nu q^\beta \, . 
\end{equation}

\paragraph*{}
By using those vertices, which include a deformation of the original QED vertex, we find  corrections for the leptonic electric and magnetic dipole moments~\cite{EDM2008, EDMMDM}. 

\paragraph*{}
Since we are interested in determining the LV effects on the elastic LbyL scattering, in the next step we shall use Eq.~\eqref{nonm1} and Eq.~\eqref{nonm2} to calculate scattering amplitudes and the associated differential cross sections. This shall be done in the sequel.

\section{Elastic Light-By-Light scattering with Lorentz symmetry violation}

\subsection{$F$-Coupling}

\paragraph*{}
The elastic LbyL scattering which we wish to analyze depends on a tensor that; at the 1-loop approximation, it corresponds to the following expression
\begin{eqnarray}\nonumber
&&T^{\mu \nu \alpha \beta}(q_1,q_2,q_3,q_4) = \int \!\! \frac{ d^d p}{(2 \pi)^d} Tr\Big{[}S(p)\gamma^\mu S(p-q_4) \times \\
&& \,\,\,\,\,\,\,\,\,\,\,\, \times \,\, \gamma^\nu S(q-q_4-q_3)\gamma^\alpha S(p-q_1) \gamma^\beta \, ,
\end{eqnarray}
where $S(p)$ is the fermion propagator with momentum $p$. The key to visualize the contribution of the LV modification is to rewrite the vertex as
\begin{eqnarray}\nonumber
\Gamma^\mu(q) &=& e \gamma^\alpha (\delta_\alpha^\mu + i e^{-1}\xi \cdot q \delta_\alpha^\mu  - i e^{-1}q_\alpha \xi^\mu) \\
&=&  e \gamma^\alpha M_\alpha^{~\mu}(\xi, q) \, ,
\label{newformnom1}
\end{eqnarray}
and, since the vertex does not depend on $p$, i.e., the internal momentum integrated in the loop, the contributions coming from the LV terms factorizes. The tensor with the LV contribution reads as follows
\begin{eqnarray}\nonumber
&& \!\!\!\!\!\!\!\!\! T^{\mu' \nu' \alpha' \beta'}_{\textmd{LV}}(q_1,q_2,q_3,q_4)= T^{\mu \nu \alpha \beta}(q_1,q_2,q_3,q_4) \times \\
&\times& \!\! M_\mu^{~\mu'}\!(\xi,q_1)M_\nu^{~\nu'}\!(\xi,q_2)M_\alpha^{~\alpha'}\!(\xi,q_3) M_\beta^{~\beta'}\!(\xi,q_4) \, .
\end{eqnarray}

\paragraph*{}
The result could also be seen from the point of view of the coupling between the tensor $T^{\mu \nu \alpha \beta}$ and the redefined polarization vectors $\epsilon_\mu(q_i)$. The LV modifies the polarization vectors themselves, i.e.,
\begin{eqnarray}\nonumber
\epsilon'_\mu(q) &=& M_\mu^{~\alpha}(q) \epsilon_\alpha(q)  \\
&=&(1+ i e^{-1}\xi \cdot q )\epsilon_\alpha(q) - i e^{-1} \xi_\mu q \cdot \epsilon(q) \, .
\end{eqnarray}

\paragraph*{}
Considering physical external photons satisfying the gauge condition $\partial_\mu A^\mu = 0$, we have $q \cdot \epsilon(q)=0$, we reach 
\begin{equation}
\epsilon'_\mu(q) = (1+ i e^{-1}\xi \cdot q )\epsilon_\alpha(q) \, .
\end{equation}

\paragraph*{}
Using the scattering matrix  
\begin{eqnarray}\nonumber
&&\mathcal{M}^{\textmd{QED}}_{\lambda_1,\lambda_2,\lambda_3,\lambda_4 } =(\epsilon_1^{\lambda_1})_\mu(\epsilon_2^{\lambda_2})_\nu(\epsilon_3^{\lambda_3})^*_\alpha (\epsilon_4^{\lambda_4})^*_\beta  \times \\
&& \,\,\,\,\,\,\,\,\,\,\,\, \times \,\, T^{\mu \nu \alpha \beta}(q_1,q_2,q_3,q_4) \, , 
    \end{eqnarray}
and taking into account the LV contribution, we arrive at the following modified scattering matrix
\begin{equation}
\mathcal{M}^{\textmd{LV}}_{\lambda_1,\lambda_2,\lambda_3,\lambda_4 } = (1 + C) \mathcal{M}^{\textmd{QED}}_{\lambda_1,\lambda_2,\lambda_3,\lambda_4 } \, ,
\end{equation}
where 
\begin{eqnarray}
C &=&i e^{-1} (q_1 +q_2-q_3-q_4)\cdot\xi  + e^{-2} [-(q_1\cdot\xi) (q_2\cdot\xi) + \nonumber \\
&+&(q_1\cdot\xi)  (q_3\cdot\xi) +(q_1\cdot\xi)  (q_4.\xi) +(q_2\cdot\xi)  (q_3\cdot\xi) +\nonumber \\
&+&(q_2\cdot\xi)  (q_4\cdot\xi) -(q_3\cdot\xi)  (q_4\cdot\xi) ]+O(\xi^3) \, .
\end{eqnarray}
Since by momentum conservation $q_1 +q_2-q_3-q_4 = 0$, the LV only contributes to second order of the $\xi$ parameter. Finally, we have $|M_{if}|^2$ given by
\begin{eqnarray}\nonumber
|\mathcal{M}^{\textmd{LV}}|^2 &=& (1+C)(1+C^*)|\mathcal{M}^{\textmd{QED}}|^2 \\
&\approx& (1+2 \textmd{Re}(C)) |\mathcal{M}_{\textmd{QED}}|^2  \, .
\end{eqnarray}
 
\paragraph*{}
Choosing a reference frame where, in the Lorentz gauge, $q_1 = (\omega,\boldsymbol{q})$, $q_2 = (\omega,-\boldsymbol{q})$, $q_3 = (\omega,\boldsymbol{k})$ and $q_4 = (\omega,-\boldsymbol{k})$ , the LV contribution can be brought into the form
\begin{eqnarray}\nonumber
C = e^{-2}\big{[}(\boldsymbol{k} \cdot \boldsymbol{\xi})^2 +(\boldsymbol{q} \cdot \boldsymbol{\xi})^2  + 2(\omega \xi_0)^2  \big{]}+ O(\xi^3) \, .
\end{eqnarray}
where $\xi^\mu = (\xi_0, \boldsymbol{\xi})$. Without loss of generality, a reference frame where the incoming photons are on the z-axis, i.e., $\boldsymbol{q}=\omega \boldsymbol{\hat{z}}$ and $\boldsymbol{k}\cdot \boldsymbol{\hat{z}}=\omega \cos \theta$ can be chosen. 
Writing $\boldsymbol{\xi}$ in an arbitrary direction 
\begin{equation}
\boldsymbol{\xi}/|\boldsymbol{\xi}| = \sin \theta_\xi \cos \phi_\xi \boldsymbol{\hat{x}} + \sin \theta_\xi \sin \phi_\xi \boldsymbol{\hat{y}} + \cos \theta_\xi  \boldsymbol{\hat{z}}\, ,
\end{equation}
we have
\begin{eqnarray}
\boldsymbol{k} \cdot \boldsymbol{\xi} &=& |\boldsymbol{\xi}|\omega \left( \sin \theta \sin \theta_\xi \cos (\phi - \phi_\xi) + \cos \theta \cos \theta_\xi\right) \nonumber \\
\boldsymbol{q} \cdot \boldsymbol{\xi} &=& |\boldsymbol{\xi}|\omega \cos \theta_\xi 
\, .
\end{eqnarray}

\paragraph*{}
Taking these relations involving $\boldsymbol{q}$, $\boldsymbol{k}$ and $\boldsymbol{\xi}$ in terms of $\omega$ and the angles $\theta$, $\phi$, $\theta_\xi$ and $\phi_\xi$, the result will be in general
\begin{eqnarray}
&&\frac{1}{4} \sum |\mathcal{M^{\textmd{LV}}}|^2 \,\approx\, \frac{1}{4} \sum  \Big{(}1 +  \omega^2 \rho^2 \Big{)}|\mathcal{M^{\textmd{QED}}}|^2 \, ,
\end{eqnarray}
where $\rho^2 = \rho^2(\theta,\phi, \theta_\xi, \phi_\xi) = 2 \textmd{Re}( C )/\omega^2$. 

\paragraph*{}
In so doing, the final result for modified differential cross section can be written as  
\begin{eqnarray}\nonumber
\frac{d \sigma^{\gamma\gamma, \xi}}{d \Omega} &=& \frac{1}{64 \pi^2} \frac{1}{(2 \omega)^2} \sum|\mathcal{M}^{\textmd{QED}}|^2  \\
&=& \frac{d \sigma^{\gamma\gamma}_{\textmd{QED}}}{d \Omega}\Big{(}1 +  \omega^2 \rho^2(\theta,\phi, \theta_\xi, \phi_\xi) \Big{)} \, .
\end{eqnarray}

\subsection{$\tilde{F}$-Coupling}

For the $\tilde{F}$-coupling, the analysis is the same as in the previous Section. However, the results differ due to the new structure brought by the Levi-Civita tensor. Rewriting the result in Eq.~\eqref{nonm2} in an analogous way to what we have done in Eq.~\eqref{newformnom1}, we have
\begin{eqnarray}
\Gamma^\mu &=&  e\gamma^\alpha(\delta^\mu_\alpha - e^{-1} \varepsilon^\mu_{~  \alpha \nu \beta}\tilde{\xi}^\nu q^\beta) \nonumber \\
&=& e \gamma^\alpha N^\mu_{~\alpha} (\tilde{\xi}, q)  \, .
\label{newformnom2}
\end{eqnarray}

\paragraph*{}
In order to calculate the differential cross section for elastic LbyL scattering with this kind of LV term, we rewrite the polarization vector. But, in this case, we use the term $N_{\alpha}^\beta$, so that
\begin{eqnarray}
\epsilon^{'\mu}(q) &=& N^{\mu}_{~\nu}(q) \epsilon^\nu(q) \nonumber \\ &=&\epsilon^\mu(q) - i e^{-1}\varepsilon^\mu_{~  \nu \alpha \beta}\tilde{\xi}^\alpha q^\beta \epsilon^{\nu}(q) \, ,
\end{eqnarray}
and, in the same way we have done for the $F$-coupling, we reach a new scattering matrix as follows
\begin{eqnarray}\nonumber
&&\mathcal{M}^{\textmd{LV}}_{\lambda_1,\lambda_2,\lambda_3,\lambda_4 } = (\epsilon_1^{\lambda_1})^\mu(\epsilon_2^{\lambda_2})^\nu(\epsilon_3^{\lambda_3})^{*\alpha} (\epsilon_4^{\lambda_4})^{*\beta}   T_{\mu \nu \alpha \beta}  \\
\nonumber && ~~~~~~~~~~~~ = N^{\mu}_{~\mu'}  N^{\nu}_{~\nu'} (N^*)^{\alpha}_{~\alpha'} (N^*)^{\beta}_{~\beta'} \times\\ && ~~~~~~~~~~ \times ~  (\epsilon^{\lambda_1})^{\mu'}(\epsilon^{\lambda_2})^{\nu'}(\epsilon^{\lambda_3})^{*\alpha'} (\epsilon^{\lambda_4})^{*\beta'} T_{\mu \nu \alpha \beta} \, .
\end{eqnarray}
So, in a short way,
\begin{eqnarray}\nonumber
|\mathcal{M}^{\textmd{LV}}|^2 &=& (NN)_{~\mu''}^\mu(NN)^\nu_{\nu''}(NN)^\alpha_{\alpha''}(NN)^\beta_{\beta''}\times\\
&&~~~~\times T_{\mu \nu \alpha \beta} (T^*)^{\mu'' \nu'' \alpha'' \beta''} \, ,
\end{eqnarray}
where $(NN)_{~\mu''}^\mu =N^{\mu}_{~\mu'} N^{*\mu'}_{\mu''} $ and the contractions in parenthesis can be expanded as
\begin{equation}
(NN)_{~\mu''}^\mu \!= \delta^\mu_{\mu''}(1-(q \cdot \tilde{\xi})^2)  + q^2( \delta^{\mu'}_{\mu''} \tilde{\xi}^2 - \tilde{\xi}^\mu \tilde{\xi}_{\mu''}) + O(\tilde{\xi}^3) .
\end{equation}

\paragraph*{} 
Here, we ignore terms proportional to $q_\mu$, which cancel contracted with any index of $T$ due to gauge invariance. Finally, we can find $|M|^2$ and the differential cross section. One more time, considering external physical photons, we have 
\begin{equation}
\label{Fdualresult}
\frac{d \sigma^{\gamma\gamma, \tilde{\xi}}}{d \Omega} = \frac{d \sigma^{\gamma\gamma}_{\textmd{QED}}}{d \Omega} \Big{(}1 - \omega^2\rho^2(\theta,\phi, \theta_{\tilde{\xi}}, \phi_{\tilde{\xi}}) \Big{)} \, ,
\end{equation}
where $\rho$ is the same function found for the differential cross section in the case of the background 4-vector $\xi$.

\paragraph*{}
In both nonminimal couplings, the LV contributions show up in regions where $\omega_\xi \approx e/|\rho| $. Since the LV parameters come typically from high-energy effects, an analysis in the UV limit could be more productive. 

\paragraph*{}
The modified vertices that we derive for the $F$- and $\tilde{F}$-couplings manifest themselves as nonrenormalizable, since, in this process, they are internal vertices in Feynman diagrams. However, if LV effects are observed on the energy scale of the currently available experiments, we expect them to be manifestations of a more fundamental theory at high energies. So, when we operate with this theory below its characteristic mass scale, it is acceptable to work with a nonrenormalizable model, taking the standpoint that it is seen as an effective field theory valid below a certain cutoff \cite{effect1}. In this work, even when we consider elastic LbyL scattering in the high-energy regime, we are still considering energies far below the characteristic mass scale of the theory. And so, the analysis performed here is meaningful in the energy region we are concerned with.

\paragraph*{}
Now, to shed light on our results, some particular cases will be discussed and we shall show as different particularizations considered on $\xi^\mu$ modify the differential cross section of the elastic LbyL scattering.

\section{Differential Cross Section: LV Effects}

\paragraph*{}
In order to visualize the LV effects on the differential cross section of elastic LbyL scattering, we take some particular choices for the configuration of the background 4-vector. Since the extra term $\omega^2 \rho^2$ is common to both $F$- and $\tilde{F}$-couplings, with only a signal difference, we shall only analyze the former. 

The first choice is a timelike and the second one a spacelike $\xi^\mu$. We could also consider a lightlike 4-vector, i.e., $\xi = (\zeta,0,0,\zeta)$. However, this third case presents a superposition of the effects from the first two cases, so we omit it. In practice, such divisions are arbitrary: if $\xi$ exists, it will be a nontrivial mixture of temporal and spacial components.

\subsection{Timelike background 4-vector}

\paragraph*{}
In the case of a timelike LV, i.e., $\xi^\mu = (\xi_0, \boldsymbol{0})$, the result takes a simple form. The contribution will be given by $\rho^2(\theta, \phi, \theta_\xi, \phi_\xi) = 4 e^{-2} \xi_0^2$ and consequently
\begin{equation}
\Big{|} ~\frac{\frac{d \sigma_{_{\textmd{QED}}}}{d \Omega}}{\frac{d \sigma}{d \Omega}} -1 ~\Big{|} \approx 4 e^{-2} \xi_0^2 \omega^2
\end{equation}

\paragraph*{}
The effects of the timelike LV will show up in frequencies near to $\omega_\xi = e \xi_0^{-1}$. 
\begin{figure}[H] 
	\begin{center}
		\leavevmode
		\includegraphics[width=0.5\textwidth]{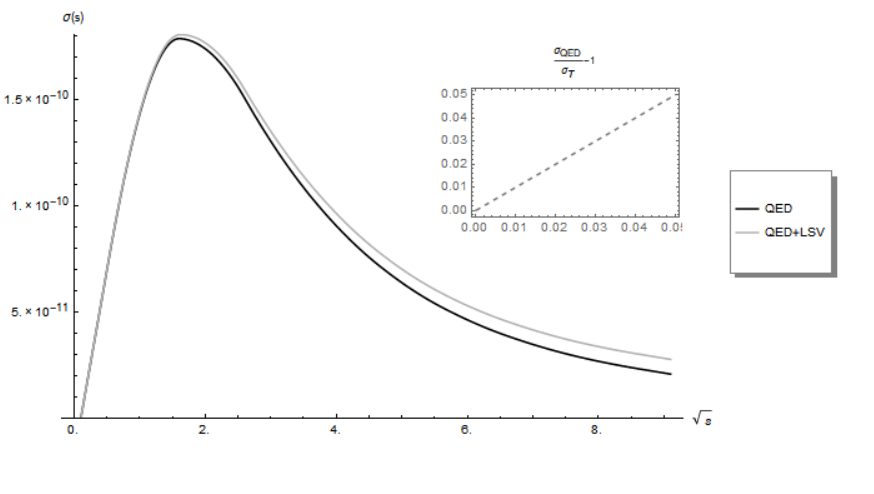}
	\end{center}
	\caption[]{Photon-photon cross section profile for QED (black) and for QED plus time-like Lorentz violation contribution with $\omega_\xi = 0.01 m$ (gray).}
	\label{crtl}
\end{figure}
\paragraph*{}
Since the difference between the QED cross section and the one modified by LV terms increases with the photon energy, the best regime to detect possible LV signals is in the high-energy limit, in which the QED contribution turns down and the LV effect rises up. 

\paragraph*{}
By introducing cutoff parameters, $\Lambda_\pm$, in a modified differential cross section of the process \cite{Derrick86}, and by comparison between the new differential cross section and experimental results, it is possible to parameterize deviations from QED. As soon as experimental data for elastic LbyL scattering are available with good precision, this analysis can be adopted to find the values of $\Lambda_\pm$ with an adequate confidence level. Finally, we get upper bounds for our LV parameters in the same spirit as it has already been done for other QED processes \cite{LSVbhabha, LSVsaladirac}.

\subsection{Spacelike background 4-vector}

\paragraph*{}
We next consider the case of a purely spacelike background 4-vector, i.e., $\xi^\mu = (0, \boldsymbol{\xi})$. In this particular case, we shall investigate the angular profile of differential cross section in the low- and high-energy regimes for elastic LbyL scattering.

\paragraph*{}
With the choice of reference frame already adopted in the previous Section, with $\boldsymbol{\xi}$ in an arbitrary direction, we can split the total differential cross section as a part due to only QED and another part due to LV, i.e.
\begin{equation}
\frac{d\sigma^{\gamma \gamma , \boldsymbol{\xi}}}{d\Omega} = \frac{d\sigma^{\gamma \gamma}}{d\Omega}+\frac{d\sigma^{\gamma \gamma, \boldsymbol{\xi}}_{_{\textmd{LV}}}}{d\Omega} \, .
\end{equation}

\paragraph*{}
We then verify the effects of the proposed modification on the differential cross sections, both in the low- and in the high-energy regime. If $\omega \ll m$, a general result for the spacelike 4-vector case is written as
\begin{eqnarray}
\! \!\frac{d\sigma^{\gamma \gamma, \boldsymbol{\xi}}_{_{\textmd{LV}}}}{d\Omega} \! \! &=& \frac{2|\boldsymbol{\xi}|^2}{e^2} \frac{139 \alpha^4 \omega^8}{(180 \pi)^2m^8} (3 + \cos^2\theta)^2 \Big{[} \left( \cos \theta_\xi \right)^2+  \nonumber \\
 &+& \!\! \left( \sin \theta \sin \theta_\xi \cos (\phi - \phi_\xi) + \cos \theta \cos \theta_\xi \right)^2  \! \Big] \!,
\end{eqnarray}
which offers a better visualization of the LV effects in the two energy regimes. 

\paragraph*{}
Taking first a background vector parallel to the z axis $(\theta_\xi = 0)$, we find
\begin{equation}
\! \!\frac{d\sigma^{\gamma \gamma, \parallel}_{_{\textmd{LV}}}}{d\Omega} \! \! = \frac{2|\boldsymbol{\xi}|^2}{e^2} \frac{139 \alpha^4 \omega^8}{(180 \pi)^2m^8} \Big[ 9 + 15\cos^2\theta + 7\cos^4\theta+ \cos^6\theta \Big]  ,
\end{equation}
from which we can see that there is a change in the angular dependence on $\theta$ and no azimuthal dependence. 

\begin{figure}[h]
	\begin{center}
		\leavevmode
		\includegraphics[width=0.4\textwidth]{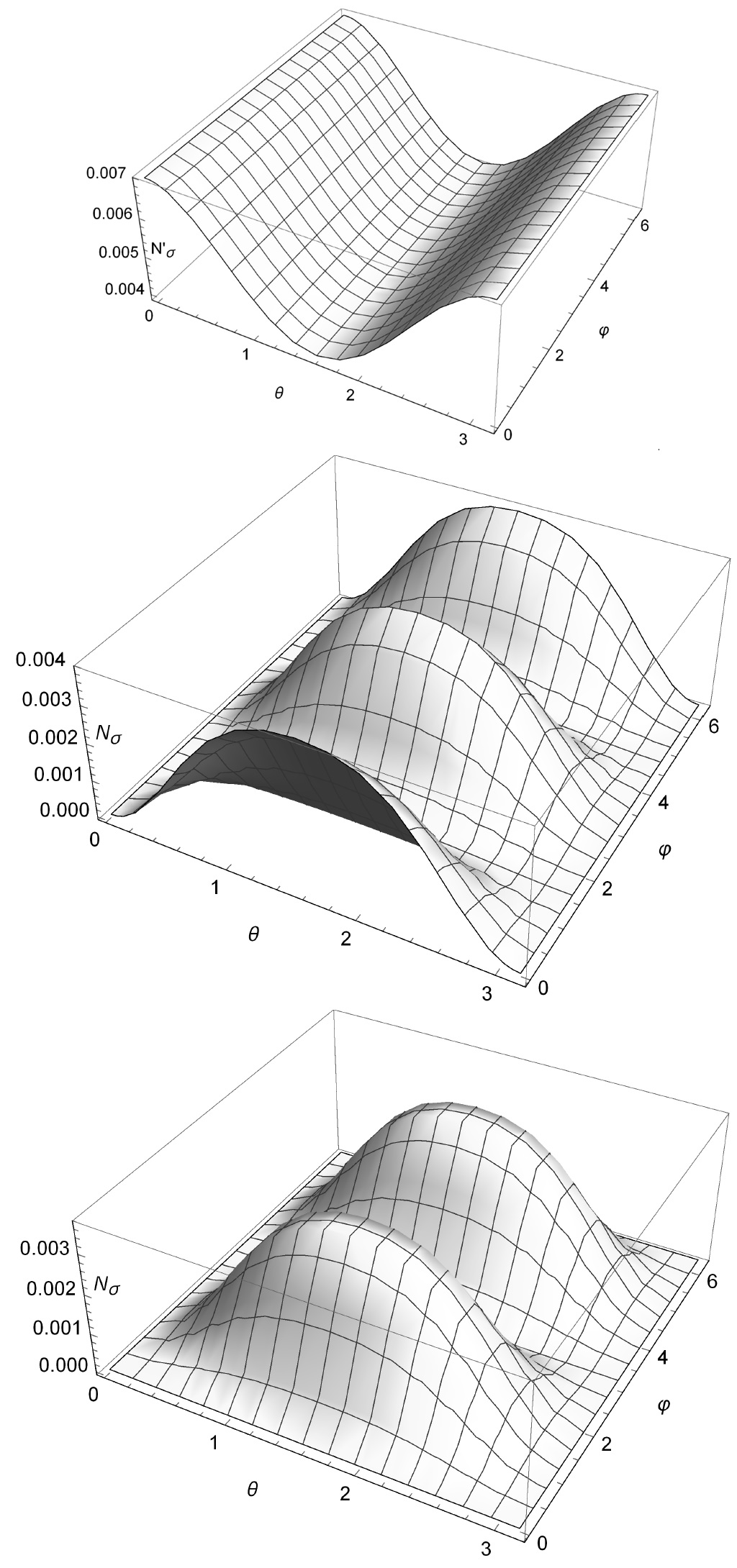}
	\end{center}
	\caption[QED em Baixas Energias e LV]{Instantaneous angular profile of differential cross sections for QED (top) and purely spacelike background $(\boldsymbol{\xi} \perp \boldsymbol{\hat{z}}, i.e., \theta_\xi=0)$ LV scenario, in low-energy regime. The vertical axes are given by $N'_\sigma = [ \alpha^4 \omega^6 / m^8 ]^{-1} d\sigma^{\gamma \gamma}/d\Omega$ and $N_\sigma = [ 2\alpha^4 |\boldsymbol{\xi}|^2\omega^8 / e^2 m^8 ]^{-1} d\sigma^{\gamma \gamma,\perp}_{_{\textmd{LV}}}/d\Omega$, with $\phi_\xi=0$ (middle) and $\phi_\xi=\phi/2$ (bottom). }
	\label{LVGraflow}
\end{figure}

\paragraph*{}
On the other hand, if we consider the background vector on the transverse plane xy $(\theta_\xi = \pi/2)$, we have
\begin{equation}
\! \!\frac{d\sigma^{\gamma \gamma, \perp}_{_{\textmd{LV}}}}{d\Omega} \! \! = \! \frac{2|\boldsymbol{\xi}|^2}{e^2} \frac{139 \alpha^4 \omega^8}{(180 \pi)^2m^8} [ \left( 3 + \cos^2 \! \theta \right) \! \left( \sin \! \theta  \cos (\phi - \phi_\xi)\right)  ]^2  \! ,
\end{equation}
whose angular profile is plotted in Fig.~\ref{LVGraflow} for different relative orientations of the background vector in the transverse xy plane. Here, we have a clear $\phi$-dependent feature that is very distinctive relative to the profile from QED. This distinctive angular dependence could be used as am experimental signature to be searched for.

\paragraph*{}
Now, we turn into the high-energy regime, under the conditions of validity of Eq. \eqref{eqHighRegime}. Similarly to the previous one, taking a background vector parallel to the z axis we obtain
\begin{equation}
\label{highLVparallel}
\! \!\frac{d\sigma^{\gamma \gamma, \parallel}_{_{\textmd{LV}}}}{d\Omega} \! \! = \frac{2|\boldsymbol{\xi}|^2}{e^2} \frac{\alpha^4}{\pi^2 } \log^4 \frac{1}{\theta} \Big[ 1 + \cos^2\theta  \Big] 
\end{equation}
and again, more interesting is the second scenario with a transverse background vector, in which the LV piece becomes
\begin{equation}
\label{highLVperp}
\! \!\frac{d\sigma^{\gamma \gamma, \perp}_{_{\textmd{LV}}}}{d\Omega} \! \! = \frac{2|\boldsymbol{\xi}|^2}{e^2} \frac{\alpha^4}{\pi^2 } \log^4 \frac{1}{\theta} \left( \sin^2 \theta  \cos^2 (\phi - \phi_\xi)\right)   .
\end{equation}

\begin{figure}[h]
	\begin{center}
		\leavevmode
		\includegraphics[width=0.4\textwidth]{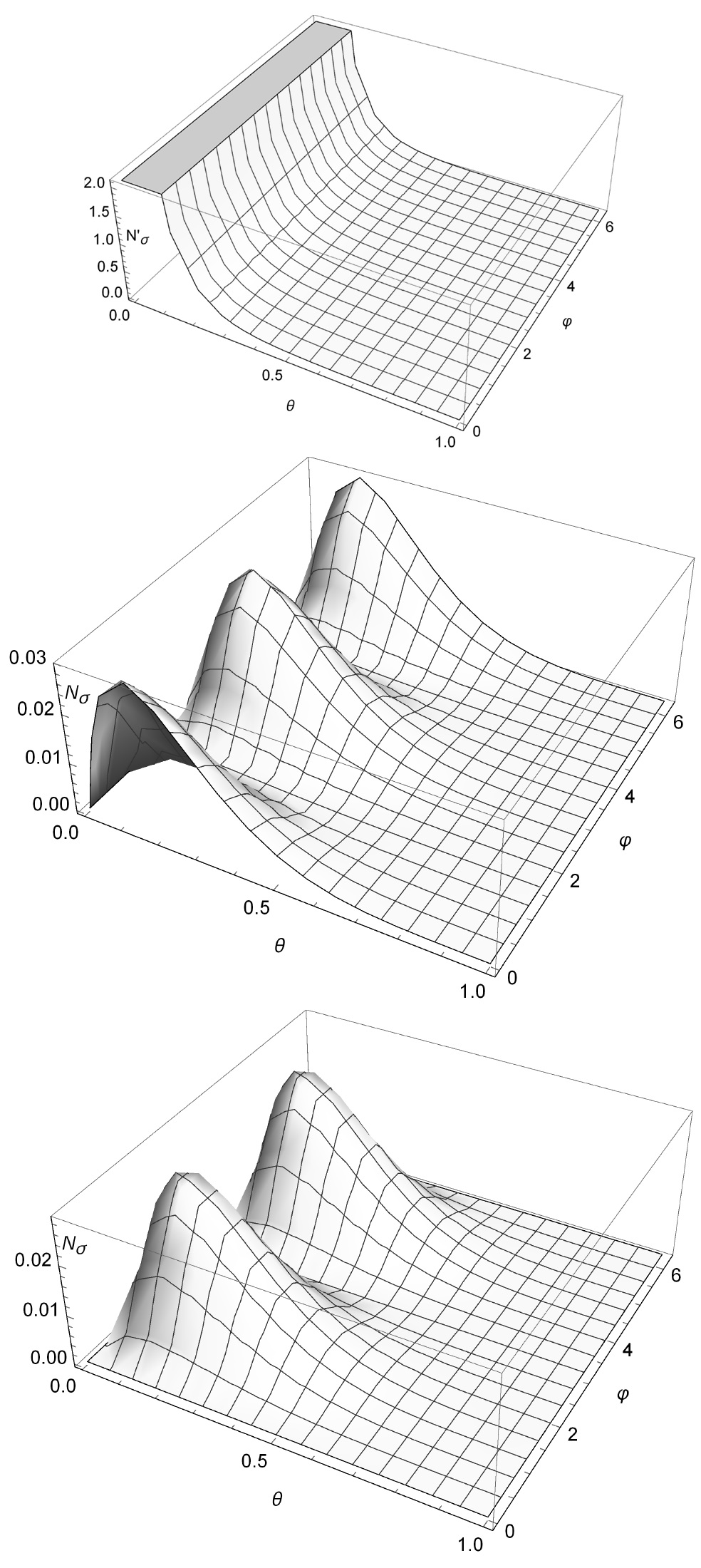}
	\end{center}
	\caption[QED em Altas Energias e LV]{Instantaneous angular profile of differential cross sections for QED (top) and purely spacelike background $(\boldsymbol{\xi} \perp \boldsymbol{\hat{z}}, i.e., \theta_\xi=0)$ LV scenario, in high-energy regime. The vertical axes are given by $N'_\sigma = [ \alpha^4 / \omega^2 ]^{-1} d\sigma^{\gamma \gamma}/d\Omega$ and $N_\sigma = [ 2\alpha^4 |\boldsymbol{\xi}|^2/ e^2]^{-1} d\sigma^{\gamma \gamma,\perp}_{_{\textmd{LV}}}/d\Omega$, with $\phi_\xi=0$ (middle) and $\phi_\xi=\phi/2$ (bottom). }
	\label{LVGrafhigh}
\end{figure}

\paragraph*{}
This LV contribution is also plotted in Fig. \ref{LVGrafhigh} for different choices of the azimuthal angle, $\phi_\xi$. One more time, we have a anisotropic profile in the scattering cross section.  

\paragraph*{}
When we analyze the cases of differential cross sections modified by a 4-vector background on the transverse plane, the angular profile in low-energy regime for QED graphs shows minimum values in $\theta=\pi/2$ while the LV surfaces has maxima in the same $\theta=\pi/2$. These maxima occur in $\phi=0, \pi, 2\pi$ for $\boldsymbol{\xi} \parallel \boldsymbol{\hat{z}}$ and in $\phi = \pi/2, 3\pi/2$ for $\boldsymbol{\xi} \perp \boldsymbol{\hat{z}}$. Such specifications are those in which it would be easier to observe LV effects in experiments.

\paragraph*{}
In the high-energy regime, the LV effects will be more salient for the same $\phi=0, \pi, 2\pi$ for $\boldsymbol{\xi} \parallel \boldsymbol{\hat{z}}$ and in $\phi = \pi/2, 3\pi/2$ for $\boldsymbol{\xi} \perp \boldsymbol{\hat{z}}$, but the smaller the value of $\theta$ the more intense it will be the LV effects. 

\paragraph*{}
Furthermore, Eqs. \eqref{highLVparallel} and \eqref{highLVperp} show that the extra LV contribution up to $O(\xi^3)$ is energy-independent, while the differential cross section of QED falls off with $\omega^{-2}$. In the case $\boldsymbol{\xi} \perp \boldsymbol{\hat{z}}$, in particular, a plateau could be observed with experiments performed at increasingly higher energies for small $\theta$. It is important to remember that the limits of validity of Eq. \eqref{eqHighRegime} ($ m/ \omega \! \ll \! \theta \! \ll \! 1$) must be respected.

\paragraph*{}
If the whole previous analysis were developed for the $\tilde{F}$-coupling, the same resulting azimuthal dependence would be observed, but with a global minus sign, as indicated in Eq. \eqref{Fdualresult}. Similar results have been reported with this $\phi$-dependence, considering nonminimal couplings to modify the QED Lagrangian, in processes such as Compton and Bhabha scatterings \cite{LSVbhabha,LSVsaladirac}.

\paragraph*{}
The above LV considerations apply for a truly fixed and time-independent 4-vector background. These requirements are only explicit in an inertial frame. Due to Earth motion, we know that this is not the case for laboratory frames. Therefore, in considering those, the background should vary over time. An approximately inertial frame commonly used in the literature is the so-called Sun-centered frame (SCF) \cite{Mewes}.

\paragraph*{}
In order to describe the 4-vector observed in the Earth reference frame, $ \xi_{\rm lab}$, from $ \xi_{\rm Sun}$, we use a Lorentz transformation, i.e., $\xi_{\rm lab}^{\mu} = \Lambda^{\mu}_{\, \, \, \nu} \, \xi_{\rm Sun}^{\nu}$, where $\Lambda^{\mu}_{\, \, \, \nu}$ can be seen in ref. \cite{Mewes}. Since the relativistic effects, quantified by $\beta << 1 $, are small, we can write 
approximately $\xi^{0}_{\rm lab} = \xi^{T}_{\rm Sun} \equiv 0$ and $\xi^{i}_{\rm lab} = R^{i \, j} (\chi, T_{\oplus})\,  \xi^{j}_{\rm Sun}$ where the rotation matrix depends on the time in the SCF $T_{\oplus}$.

\paragraph*{}
Since experiments are usually driven by long time scales, the LV signatures in reference frames on Earth should correspond to the temporal average of these effects. Therefore, $\xi_{\rm lab}^{x} = -\sin\chi \, \xi_{\rm Sun}^{z}$ and $\xi_{\rm lab}^{z} = \cos\chi \, \xi_{\rm Sun}^{z}$ will be the non-vanishing spatial components, where $\chi$ 
is the colatitude of the laboratory. Thus, the vector components $ \xi_{\rm lab} $ would be related with the background in the SCF for any experiment.

\section{Concluding Remarks}

\paragraph*{}
In this contribution, we have investigated two specific nonminimal Lorentz-violating couplings between the fermion and gauge fields, and their effects on the differential cross section of elastic photon-photon scattering as described by QED. When considering the coupling with $F$, a new $\boldsymbol{\xi}$-dependent charge conservation scheme was derived. Nevertheless, the usual electric charge conservation is not affected. In the special case where we consider only a timelike background
4-vector, this $\xi$-dependent contribution to the new charge vanishes. The coupling with $\tilde{F}$ generates the same kind of contribution to the differential cross section of elastic LbyL scattering without modification in the charge conservation law, even in the spacelike case.

\paragraph*{}
Another advantage we can be point out is that these couplings yield electric and magnetic transition dipole moments, which can open up the opportunity to find upper bounds for the parameters from processes involving neutrinos, for example.

\paragraph*{}
In our development, we have not presented results dependent on a general $\xi$, with both temporal and spatial components. Rather, we have focused on purely timelike or spacelike LV backgrounds. This choice -- common in works on LV -- gives us differential cross section modifications which can be more easily observed. With new experimental results, it might be possible to estimate upper bounds for the LV parameters of our models. Since the data stemming from experiments involving ultraperipheral Pb-Pb collisions at LHC are expected to increase tenfold after LHC Run 4~\cite{LbyLCERN2017}, scheduled to start in 2026, we expect to have accurate data to find these upper bounds in the next decade.

\paragraph*{}
When we analyze the behavior of the differential cross section in the spacelike case, we observe that new terms bring extra angle dependences, both in $\theta$ and $\phi$. The $\phi$-contribution is the most interesting one, since the QED results are independent of azimuthal variation, and, in our case, this is no longer necessarily true. There arises a periodic pattern in $\phi$ angle in the high- and low-energy limits (Fig. \eqref{LVGraflow} and Fig. \eqref{LVGrafhigh}), and this could be a visible sign of the Lorentz violation from an experimental point of view. This azimuthal dependence is present in other QED+LV processes, as Compton, Bhabha and M{\o}ller scatterings~\cite{LSVbhabha,LSVsaladirac} considering the couplings in Eqs.~\eqref{nonm1} and~\eqref{nonm22}. Besides that, since that the QED contribution to photon-photon scattering decreases in the high-energy limit, whereas the LV contribution increases, this regime would be the most fruitful regime to search for LV effects.

\paragraph*{}
One remark is in order: the LV couplings we discussed are inspired by the CFJ model~\citep{YM}, but here the charged current replaces the photon field. In this scenario, we consider that the effects of LV from our nonminimal coupling should be more easily observed than in the CFJ model. Instead of modifying the propagator of the photon field, these nonminimal couplings modify the vertex of the interactions between the photon and the fermionic current. Therefore, this approach maintains the dispersion relations of the QED massless photon. Furthermore, derivative couplings naturally show up in high-energy limits; thus, the nonminimal couplings we use could be observed in high-energy experiments, as the LHC, more easily than the minimal LV coupling.

\section*{ACKNOWLEDGMENTS}
The authors are grateful to J. A. Helay\"el-Neto, V. A. Kosteleck\'y, P. C. Malta and G. A. Moth\'e for interesting discussions, helpful suggestions and for reading the manuscript. This work was funded by the Brazilian National Council for Scientific and Technological Development (CNPq). 


\end{document}